\documentclass[aps,prl, twocolumn, floatfix, showpacs]{revtex4}

\pdfoutput=1

\usepackage{graphicx}
\usepackage{latexsym}
\usepackage{amsmath}
\usepackage{graphics}
\usepackage{amssymb}
\usepackage{layout}
\usepackage{verbatim}
\usepackage{amsfonts,epsfig}
\usepackage{bm}
\usepackage{amsmath}
\newcommand{\ket}[1]{|#1\rangle}
\newcommand{\bra}[1]{\langle #1 |}

\begin{document}

\title{Majorana Fermions and a Topological Phase Transition in Semiconductor-Superconductor Heterostructures}
\author{Roman M.~Lutchyn}
\author{Jay D. Sau}
\author{S. Das Sarma}
\affiliation{Joint Quantum Institute and Condensed Matter Theory Center, Department of Physics,
University of  Maryland, College Park, Maryland 20742-4111, USA}

\date{compiled \today}
\begin{abstract}

We propose and analyze theoretically an experimental setup for detecting the elusive Majorana particle in semiconductor-superconductor heterostructures. The experimental system consists of one-dimensional semiconductor wire with strong spin-orbit Rashba interaction embedded into a superconducting quantum interference device. We show that the energy spectra of the Andreev bound states at the junction are qualitatively different in topologically trivial (i.e. not containing any Majorana) and nontrivial phases having an even and odd number of crossings at zero energy, respectively. The measurement of the supercurrent through the junction allows one to discern topologically distinct phases and observe a topological phase transition by changing the in-plane magnetic field or the gate voltage. The observation of this phase transition will be a direct demonstration of the existence of Majorana particles. 
\end{abstract}


\maketitle

The Majorana fermions were envisioned by Ettore Majorana~\cite{Majorana} in 1937 as fundamental constituents of nature. 
Majorana particles are intriguing and exotic because each Majorana particle is its own antiparticle unlike Dirac fermions where electrons and positrons (or holes) are distinct.  Recently, the search for Majorana fermions has focused on solid state systems where many-body ground states may have fundamental quasiparticle excitations which are Majorana fermions~\cite{Franz}.  Although the emergence of Majorana excitations, which are effectively fractionalized objects (ÒanyonsÓ) obeying non-Abelian anyonic statistics rather than Fermi or Bose statistics~\cite{nayak_RevModPhys'08}, in solid state systems is by itself an extraordinary phenomenon, what has attracted a great deal of attention is the possibility of carrying out fault tolerant ÔtopologicalÕ quantum computation in 2D systems using these Majorana particles~\cite{Kitaev2003}.  Such ÔtopologicalÕ quantum computation, in contrast to ordinary quantum computation, would not require any quantum error correction since the Majorana excitations are immune to local noise by virtue of their nonlocal `topological' (TP) nature~\cite{nayak_RevModPhys'08, Kitaev2003}.  The direct experimental observation of Majorana particles in solid state systems would therefore be a true breakthrough both from the perspective of fundamental physics of fractional statistics in nature and the technological perspective of building a working quantum computer.  It is therefore not surprising that there have been several recent proposals for the experimental realization of Majorana fermions (MFs) in solid state systems~\cite{dassarma_prl'05, Sau_PRL'10, Alicea_arxiv'09}.

\begin{figure}
\centering
\includegraphics[height=3in,angle=90]{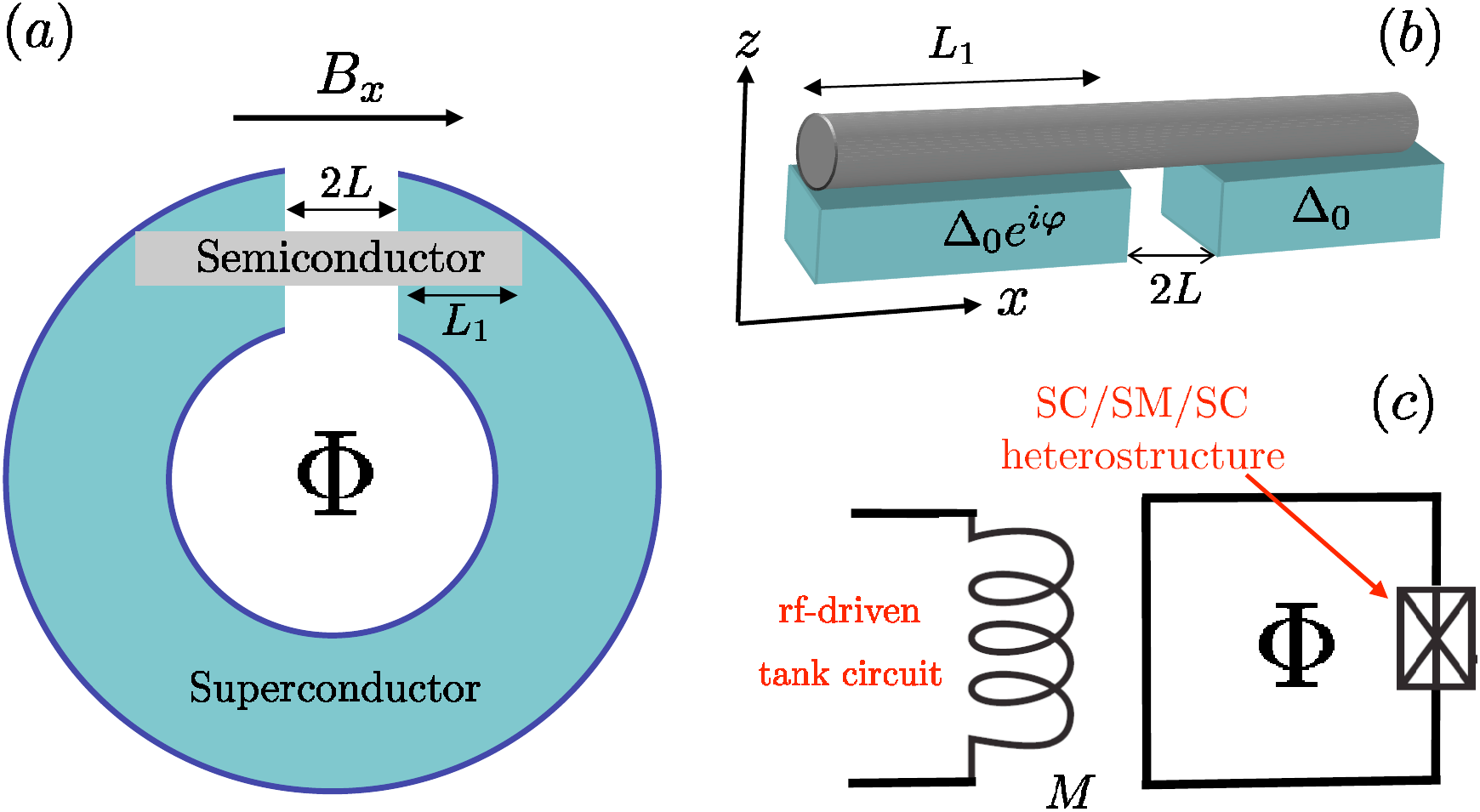}
\caption{(Color online) (a) Top view of SM/SC heterostructure embedded into small-inductance SC loop. 
(b) Side view of the SM/SC heterostructure. The nanowire can be top-gated to control chemical potential.  Here we assume $L\!\ll\! \xi$ and $L_1\!\gg\! \xi$ with $\xi$ being the SC coherence length.
(c) Proposed read-out scheme for the Andreev energy levels. Inductively coupled rf-driven tank circuit allows time-resolved measuring of the effective state-dependent Josephson inductance~\cite{Hakonen_prb'09}.}\label{fig:device}
\end{figure}

In this Letter, we propose and validate theoretically a specific experimental setup for the direct observation of MFs in one of the simplest proposed solid state systems - 1D semiconductor/superconductor heterostructure based quantum wires. This particular heterostructure consisting of an ordinary superconductor ({\it e.g.} Nb) and a semiconductor with strong spin-orbit coupling ({\it e.g.} InAs) as proposed originally by Sau et al.~\cite{Sau_PRL'10} and expanded by Alicea~\cite{Alicea_arxiv'09},  is simple and does not require any specialized materials for producing Majorana modes. The superconductor (SC) induces superconductivity in the semiconductor (SM) where the presence of spin-orbit coupling leads to the existence of MFs at the ends of the wire. 
We show that in a suitable geometry (see Fig.\ref{fig:device}) the SC state in the semiconductor undergoes a phase transition, as the chemical potential or magnetic field is tuned, from a superconducting state 
containing Majorana modes at the junction to an ordinary 
SC state with no Majorana modes at the junction.  We establish that such a transition is indeed feasible to observe in the laboratory in semiconductor nanowires, showing in the process how one can experimentally discover the Majorana mode in the SM/SC heterostructure.

Specifically, we consider here 1D InAs nanowire proximity-coupled with an s-wave superconductor ({\it e.g.}, Nb or Al).
InAs nanowires in proximity to Nb and Al have been studied experimentally\cite{Doh'05} and are known to form highly transparent interfaces for electrons 
allowing one to induce a large SC gap  $\Delta_0$ in InAs ($\Delta_0\! \lesssim \! \Delta_{\rm Nb} \!\! \approx \!\! 15$K)~\cite{Chrestin_prb'97}. Moreover, in this quasi 1D geometry
(see Fig.\ref{fig:device}b) the in-plane magnetic field $B_x$ can open up a gap in the spectrum at zero momentum and eliminate fermion doubling. Because of the vast difference in the g-factor for Nb $g_{\rm Nb}\!\sim\!1$ and InAs $g_{\rm InAs}\!\! \lesssim \!\! 35$~\cite{g-factor_InAs}, the in-plane magnetic field $B_x \!\lesssim\! 0.1$T can open a sizable Zeeman gap in InAs ($V_x \!\lesssim \! 1$K) without substantially suppressing SC in Nb ($H^{\rm Nb}_{c} \!\! \sim \!\! 0.2$T). The nanowire can be gated~\cite{Doh'05} allowing one to control chemical potential in it.
Thus, the current proposal involves a simple architecture and yet preserves the parameter phase space flexibility, which puts the realization of MFs in the SM/SC heterostructure within the experimental reach.

We show below that the supercurrent through SM/SC heterostructure exhibits unusual behavior due to the presence of MFs in the system. In particular, the spectrum of Andreev states has an odd number of crossings at $E\!=\!0$ in the TP phase ($C_0\!\!\equiv\! \mu^2\!+\!\Delta_0^2\!-\!V_x^2\!\!<\!\!0$ with $\mu$ being chemical potential)  whereas in the TP trivial phase ($C_0\!\!>\!\!0$) the number of crossings is even. 
Odd number of crossings is associated with the presence of MFs in the system leading to $4\pi$-periodic Andreev energy spectrum~\cite{kitaev'01}.
Thus, this difference in the spectrum allows distinguishing TP and conventional SCs. 
The remarkable feature of the present proposal is that by changing $B_x$ or $\mu$ across the phase boundary between TP trivial and nontrivial superconducting phases  ($C_0\!=\!0$)  one can contrast different qualitative dependence of the Andreev energy spectrum on magnetic flux $\Phi$ through the SQUID.

{\it Theoretical model.} We consider an infinite ($L_1\!\gg\!\xi$) 1D semiconducting wire embedded into SQUID, see Fig.~\ref{fig:device}a. The Hamiltonian describing the nanowire reads ($\hbar=1$)
\begin{align}
\!\!\!H_0\!=\!\!\!\int_{-\infty}^{\infty} \!\!\!\!\!\!d x \psi_{\sigma}^\dag(x)\!\!\left(\!-\!\frac{\partial_x^2}{2m^*}\!-\!\mu\!+\!i\alpha \sigma_y \partial_x\!+\!V_x\sigma_x\!\right)_{\sigma\sigma'}\!\!\!\!\!\!\psi_{\sigma'}(x),\label{eq:H0}
\end{align}
where $m^*$, $\mu$ and $\alpha$ are the effective mass, chemical potential and strength of spin-orbit Rashba interaction, respectively. In-plane magnetic field $B_x$ leads to spin splitting  $V_x\!=\!g_{\rm SM}\mu_B B_x/2$. The radius of the wire $R$ is small compared to the Fermi wavelength $R\! \lesssim\! \lambda_F$ so that there is a single 1D mode occupied.
Because of the proximity effect between SM and SC (see Fig.~\ref{fig:device}b), Cooper pairs can tunnel into the nanowire. These correlations can be described by
$H_{\rm SC}\!=\!\int_{-\infty}^{\infty} \! dx \left(\Delta(x) \psi^\dag_{\uparrow}(x)  \psi^\dag_{\downarrow}(x)\! +\!h.c. \right).$
Here $\Delta(x)$ is the induced pairing potential in the nanowire $\Delta(x)\!=\!\Delta_0 \Theta(x\!-\!L)\!+\!\Delta_0e^{i\varphi} \Theta(\!-\!x\!-\!L)$ with $\varphi$ being the phase of the order parameter.

One can recast the full Hamiltonian $H\!=\!H_0\!+\!H_{\rm SC}$ in the dimensionless form by introducing rescaled coordinates
$\tilde x\!\equiv\! m^*\alpha x $ and energies
$\tilde E\!\equiv\!E/m^*\alpha^2$. The BdG equations then become
$\tilde H_{\rm BdG}\Psi(\tilde x)\!=\!\tilde E \Psi(\tilde x)$.
Using the convention for Nambu spinors $\Psi(x)\!=\!(u_{\uparrow}(x),u_{\downarrow}(x),v_{\downarrow}(x),-v_{\uparrow}(x))$ the BdG Hamiltonian reads
\begin{align}
\tilde H_{BdG}&=\left(- \!\frac 1 2 \partial^2_{\tilde x} \!+\!i\sigma_y \partial_{\tilde x}\!-\!\tilde \mu \right) \tau_z\!+\! \tilde V_x\sigma_x \label{eq:BdG} \\
\!&+\!\tilde \Delta \Theta(\tilde x\!-\!\tilde L)\tau_x\!+\!\tilde \Delta \Theta(-\tilde x\!-\!\tilde L)\left(\cos \varphi \tau_x\!+\!\sin \varphi \tau_y\right)\nonumber.
\end{align}
The solution of the BdG equations supplemented with appropriate boundary conditions yields the Andreev spectrum in the junction. It is useful to solve for the energy at $\varphi\!=\!\pi$. At this point the profile of the order parameter in the limit of $L\! \! \ll \! \! \xi$ forms a domain wall,
which under certain conditions can host a pair of Majorana bound states~\cite{Sau_PRL'10}. To demonstrate this we investigate the existence of zero-energy solution by solving $\tilde H_{\rm BdG}\Psi_0(x)\!=\!0$. At $\varphi\!\!=\!\!\pi$, BdG Hamiltonian~\eqref{eq:BdG} is real and, thus, one can construct real Nambu spinors $\Psi_0(x)$. 
According to the particle-hole symmetry
if $\Psi_0(x)$ is a solution, then $\sigma_y \tau_y \Psi_0(x)$ is also a solution. This imposes the constraint on the spinor degrees of freedom: $v_{\uparrow/ \downarrow}(x)\!=\!\lambda u_{\uparrow/ \downarrow}(x)$ with $\lambda\!=\!\pm 1$.  Thus, the $4\times 4$ BdG Hamiltonian can be reduced to $2 \times 2$ matrix: 
\begin{align}
\left(\begin{array}{cc}
  -\frac{1}{2}\partial^2_{\tilde x}\!-\!\tilde\mu & V_x\!+\!\lambda \tilde\Delta(\tilde x)\!+\!\partial_{\tilde x} \\
  V_x\!-\!\lambda \tilde\Delta(\tilde x)\!-\!\partial_{\tilde x} & -\frac{1}{2}\partial^2_{\tilde x}\!-\!\tilde\mu
\end{array}\right)\!\!\left(\begin{array}{c}
                    u_{\uparrow}(\tilde x) \\
                    u_{\downarrow}(\tilde x)
                  \end{array}\right)\!=\!0
                  \label{eq:effBdG}.
\end{align}
One can seek solutions of Eq.~\eqref{eq:effBdG} in the form $u_{\uparrow/\downarrow}(\tilde x)\!\propto\! e^{z\tilde x}$ and require solutions for $x \gtrless 0$ to be normalizable. Let us concentrate on the $x\!>\!0$ case. Then, the characteristic equation for $z$ following from Eq.\eqref{eq:effBdG} reads
\begin{align}\label{eq:characteristic}
z^4\!+\!4(\tilde \mu \!+\!1)z^2\!+\!8\lambda \tilde \Delta_0 z \!+\! 4C_0\!=\!0 \mbox{ with } C_0\!=\!\tilde \mu^2\!+\!\tilde \Delta_0^2\!-\!\tilde V_x^2.
\end{align}
The roots $z_i$ of the above quartic equation with real coefficients should satisfy the following constraints:
$\prod_{i\!=\!1}^4 z_i\!=\!4C_0$ and $\sum_{i\!=\!1}^4 z_i\!=\!0$.
If all $z_i$ are real and $C_0\!>\!0$, these constraints are satisfied only when the number of solutions with ${\rm Re}[z]\!\gtrless\!0$ is the same. If Eq.\eqref{eq:characteristic} has at least one complex solution $z_1\!=\!a\!+\!ib$, then there is another solution  $z_2\!=a\!-\!ib$. Since the other two solutions are given by the quadratic equation, one can express these roots in terms of $a$ and $b$: $z_{3,4}\!=\!-\!a\pm\sqrt{a^2\!-\!4C_0/(a^2\!+\!b^2)}$. Given that $|{\rm Re} [\sqrt{a^2\!-\!4C_0/(a^2\!+\!b^2)}]|\!<\!|a|$ for $C_0\!>\!0$, there are two solutions with ${\rm Re}[z]\!\gtrless\!0$, respectively. Different values of $\lambda$ change the sign of $a$, and this conclusion is valid for both channels $\lambda\!=\!\pm\! 1$. Thus, when $C_0\!>\!0$ there are two exponentially decaying solutions for $x\!\gtrless\!0$ yielding 4 coefficients to match. Since the number of constraints (4 from boundary conditions and 1 from normalization) is larger than the number of  linearly independent coefficients, there are no zero energy solutions for $C_0\!>\!0$. On the other hand, similar analysis for $C_0\!<\!0$ always yields three roots with ${\rm Re}[z] \!<\!0$ either in $\lambda\!=\!1$ or $\lambda\!=\!-1$ channels resulting in six coefficients to match. Therefore, in this case there is a pair of zero-energy Majorana states. At $C_0\!=\!0$, there is a solution with $z\!=\!0$, which corresponds to the closing of the SC bulk excitation gap~\cite{Sau_PRL'10}. Therefore, the condition $C_0\!=\!0$ gives the phase boundary between TP trivial and nontrivial SC phases~\cite{read_prb'00}.

\begin{figure}
\centering
\includegraphics[height=3.4in,angle=90]{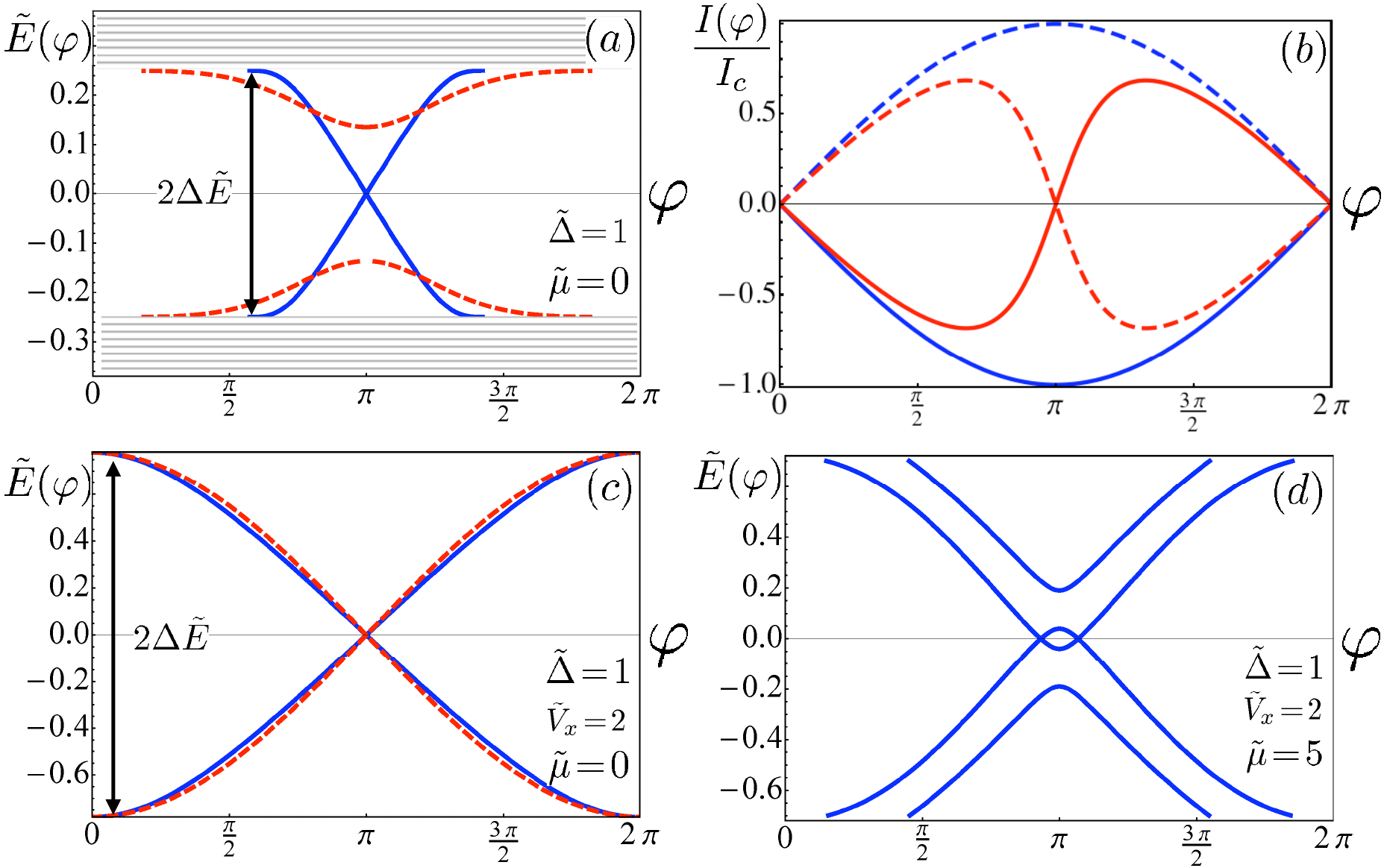}
\caption{(Color online) Andreev energy spectrum in SM/SC heterostructure for the junction with $ \tilde L  \rightarrow 0$. (a)
Energy spectrum in TP trivial (dashed line: $\tilde V_x\!=\!0.75$) and  nontrivial (solid line: $\tilde V_x\!=\!1.25$) states. 
The two TP distinct phases differ by having even and odd number of crossings, respectively.
(b) Schematic plot of the Josephson current through the junction carried by Andreev states: light (red) and dark (blue) lines describe Josephson current in TP trivial and nontrivial phases, respectively.
(c) and (d) The evolution of Andreev energy spectrum with chemical potential. (c) The spectrum in TP nontrivial phase. The dashed (red) line is a fit to $\pm \cos(\varphi/2)$ function. (d) The spectrum in TP trivial phase. There is no crossing at $\varphi=\pi$.}\label{fig:spectrumtnph}
\end{figure}

Andreev spectrum as a function of magnetic flux $\Phi$ can be obtained by solving BdG equations defined by Eq.\eqref{eq:BdG} in the limit of $L\!\!\! \rightarrow \!\!\! 0$ (describing $L\!\!\! \ll \!\!\! \xi$ case) and matching the boundary conditions $\Psi(0^{-})\!=\!\Psi(0^{+})$, $\partial_x \Psi(0^{-})\!\!=\!\!\partial_x\Psi(0^{+})$. The algebra is not particularly enlightening so we present here numerical results shown in Fig. \ref{fig:spectrumtnph}, which are consistent with above analytical considerations. The characteristic signature of the TP nontrivial phase is the presence of odd number of crossings in the Andreev spectrum in contrast with the TP trivial phase where number of crossings is even as required by $2\pi$-periodicity of the BdG Hamiltonian, see  Fig.~\ref{fig:spectrumtnph}a. Indeed, in the absence of the degenerate TP sectors, upon the advance of the SC phase $\varphi$ by $2\pi$
 the system returns to the same state. It is well-known that in SC-normal-metal-SC heterostructure the spectrum of spin-degenerate Andreev states is
$E(\varphi)\!=\!\Delta_0\sqrt{1\!-\!D\sin(\varphi/2)^2}$~\cite{Beenakker_PRL'91},
where $D$ is the interface transparency. The presence of weak spin-orbit interactions leads to the degeneracy splitting of Andreev levels~\cite{Dimitrova'06}. In the TP trivial phase, which is adiabatically connected to $V_x\!\rightarrow\! 0$ limit, we obtain similar results, see Fig. \ref{fig:spectrumtnph}d. In contrast, as shown in Fig.~\ref{fig:spectrumtnph}c Andreev spectrum in the TP nontrivial phase
is strikingly different. This difference is related to the presence of the Majorana zero-energy modes in the system at $\varphi\!=\!\pi$. The quantum phase transition between these two phases is called topological because it occurs without any qualitative changes of the local order parameter. The two phases are distinguished by the topological order associated with the presence of Majorana zero-energy modes.  The TP quantum phase transition occurs when $\Delta \tilde E$, which is proportional to the quasiparticle bulk gap, becomes zero bringing a continuum of gapless states at $\tilde E\!=\!0$. This phenomenon is generic and applies also to Majorana bound states in the vortex cores. The topological reconstruction of the fermionic  spectrum cannot occur adiabatically and requires the nullification of the bulk excitation gap~\cite{Nishida}. Looking at Figs.~\ref{fig:spectrumtnph}a and \ref{fig:spectrumtnph}c, one can see the evolution of the Andreev energy spectrum with the magnetic field: $\tilde V_x\!=\!2, 1.25, 0.75$, which supports above arguments. Also, Figs.~\ref{fig:spectrumtnph}c and \ref{fig:spectrumtnph}d show the evolution of the spectrum with the chemical potential.

We note that the position of the zero-energy crossing is not universal and can be shifted by adding a weak perturbation, e.g. $B_y \sigma_y$. However, the crossing itself is robust and is protected by particle-hole symmetry. Indeed, the eigenstates with $\pm E$ are related by particle-hole symmetry $\Psi_E\!=\!\Theta \Psi_{-E} (x)$, where $\Theta\!=\!\sigma_y\tau_y K$ with $K$ being the complex conjugation operator. One can show using the property $\sigma_y\tau_y H \sigma_y\tau_y \!=\!-H^T$ that matrix elements $\bra{\Psi} H \Theta \ket{\Psi}\!=\!-\bra{\Psi} H \Theta \ket{\Psi}\!=\!0$, and thus, the crossing is protected against any perturbations as long as the bulk gap is preserved. Another elegant way of demonstrating the robustness of the crossing point was suggested in Refs.~\cite{kitaev'01, Fu_Kane_prb'09}. At $E\!=\!0$ and $\varphi\!=\!\pi$ one can introduce two MF operators $\gamma_{1,2}\!=\!\gamma_{1,2}^\dag$. Then, the low-energy Hamiltonian around $\varphi \sim \pi$ can be written as $H\!=\!i 2 \varepsilon (\varphi) \gamma_1\gamma_2$. By introducing the Dirac fermion operators $c\!=\!\gamma_1\!+\! i \gamma_2$ and $c^\dag\!=\!\gamma_1\!-\!i\gamma_2$, one can rewrite the Hamiltonian above as $H\!=\!\varepsilon (\varphi) \! (c^\dag c\! - \!\frac{1}{2})$, from which it follows that the states $\Psi_E$ and $\Psi_{-E}$ have different fermion parity. Thus, as long as fermion parity is locally conserved the matrix elements between the states $\Psi_E$ and $\Psi_{-E}$ are zero. 

\begin{figure}
\centering
\includegraphics[height=3.4in, angle=90]{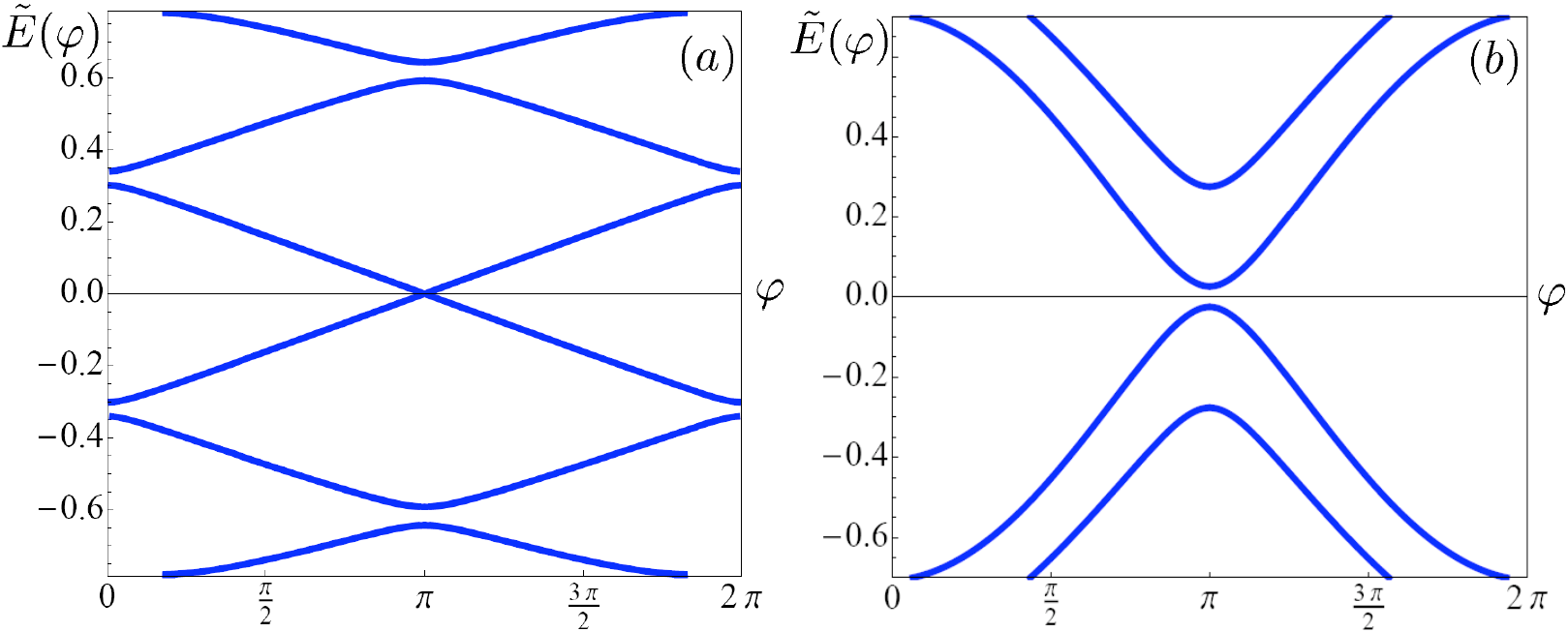}
\caption{(a) Andreev spectrum for a finite-size junction $\tilde L=3$ in a TP nontrivial phase. Here $\tilde \mu=0$, $\tilde \Delta=1$, $\tilde V_x=2$ and $U_0/\alpha=1$. (b) Andreev spectrum in TP trivial phase for $\tilde \mu=5$, $\tilde \Delta=1$, $\tilde V_x=2$, $\tilde L \ll 1$ and $U_0/\alpha=1$.} \label{fig:spectrum_imp}
\end{figure}

Two (or even number) crossings in the Andreev spectrum as in Fig.~\ref{fig:spectrumtnph}d are not generally protected. 
We have studied the robustness of even and odd crossings numerically by adding the impurity-scattering potential $U(x)=U_0\delta (x\!-\!L)$ 
into Eq.\eqref{eq:H0}. As shown in Fig.~\ref{fig:spectrum_imp}b impurity scattering opens up a gap in the spectrum indicating that crossings in the TP trivial phase are not robust. In contrast, impurity scattering does not affect the crossing in the TP nontrivial phase (see Fig.~\ref{fig:spectrum_imp}a). We also considered finite-size heterostructure $\tilde L\!=\!3$, where the spectrum has excited Andreev states. As shown in Fig.~\ref{fig:spectrum_imp}a, the crossing at zero energy is robust while other crossings are not.

The experimental system shown in Fig.~\ref{fig:device}b can be viewed as two Majorana quantum wires~\cite{kitaev'01} coupled by tunneling through the junction. Indeed, consider a SM wire of length $L_1$
at $x\!\!>\!\!0$. 
One can diagonalize the single-particle Hamiltonian~\eqref{eq:H0} and find eigenvalues $\varepsilon_{\pm}(p)\!=\!p_x^2/2m^*\!-\!\mu\pm\sqrt{V_x^2\!+\!\alpha^2 p_x^2}$ and eigenvectors $\phi_{\pm}(p)\!=\!\frac{1}{\sqrt{2}}(\pm (V_x\!+\!i \alpha p_x)/\sqrt{V_x^2\!+\!\alpha^2p_x^2},1)^T$. Assuming that only lowest band $\varepsilon_{-}(p)$ is occupied, the full Hamiltonian $H$ can be projected to the lowest band yielding
$$\!H_P\!=\!\sum_p [\varepsilon_{-}(p) c^\dag_{-}(p)c_{-}(p)\!+\!\Delta_{-}(p) c^\dag_{-}(p)c^\dag_{-}(-p)\!+\!{\rm H.c.}],\nonumber$$
where the order parameter $\Delta_{-}(p)\!=\! i \alpha p_x \Delta_0/\sqrt{\alpha^2 p_x^2\!+\!V_x^2}$ has $p$-wave symmetry. Thus, the present problem is isomorphic to the Majorana wire considered by Kitaev~\cite{kitaev'01}. As long as $L_1 \! \gg\! \xi$, tunneling amplitude of MFs between the ends of the wire vanishes $t\!\! \propto \!\! e^{-\frac{L_1}{\xi}}$ and different fermion parity ground states are almost degenerate. When two Majorana wires are brought together as shown in Fig.~\ref{fig:device}b, zero-energy Majorana modes at the junction are hybridized yielding the spectrum shown in Fig.~\ref{fig:spectrumtnph}c. The presence of MFs in the system can be characterized by $Z_2$ topological invariant $M(H_0)\!=\!(-1)^{\nu(0)-\nu(\Lambda)}$~\cite{kitaev'01}, where $\nu(0)$ and $\nu(\Lambda)$ are the number of negative eigenvalues of $H_0$ at $p\!=\!0,\Lambda$, respectively. Here $\Lambda$ is the momentum at the edge of the Brillouin zone. The difference $\nu(0)\!-\nu(\Lambda)$ counts the number of bands ($\rm mod$ 2) crossing Fermi level on the interval $(0,\Lambda)$. In weak pairing limit $\Delta_0\!\ll \! \mu, V_x$, this definition of TP trivial ($M\!=\!+\!1$) and nontrivial phases ($M\!=\!-1$) is consistent with exact results for this model discussed after Eq.~\eqref{eq:characteristic}.

The difference in Andreev spectrum should be detectable by various experimental techniques. In particular, the Josephson current in $L\! \rightarrow \! 0$ limit
 is given by
$I_n\!=\!-\!\frac{2e}{\hbar}\!\frac{\partial E_n(\varphi)}{\partial \varphi}$~\cite{Beenakker_PRL'91, Dimitrova'06}.
The energy $E_{1,2}(\varphi)$ close to $\varphi\!=\!\pi$ is well approximated by $\!\pm\! \cos(\varphi/2)$. Thus, the current carried by the quasiparticle state $n$ at $\varphi\!=\!\pi$ is maximum in the TP nontrivial phase in contrast to the TP trivial case where $I_{n}\!=\!0$, see Fig.~\ref{fig:spectrumtnph}b. This phenomenon was dubbed fractional Josephson effect~\cite{kitaev'01, Kwon'03,  Fu_Kane_prb'09}. 
In reality, however, there are processes changing fermion parity, and current will fluctuate between $I_{\pm}\!=\!\pm I$ with switching time $\tau$. Such processes were studied in the context of SC qubits ~\cite{martinis_prl'09}, where 
the fermion parity switching time $\tau$ was measured experimentally yielding $\tau\! >\!1 ms$ at $T\!=\!20$mK in Al. 
The random telegraph signal of Josephson current can be measured by inductively coupling the SQUID to the rf-driven tank circuit (see Fig.~\ref{fig:device}c) and monitoring in real time the impedance of the circuit, which depends on effective Josephson inductance $L^{-1}_J\!(\varphi)\!\!=\!\!\frac{4\pi^2}{\Phi_0^2}\frac{\partial^2 E(\varphi)}{\partial \varphi^2}$~\cite{Hakonen_prb'09}. For typical parameters of InAs $m^*\!\approx \! 0.04 m_e$, $\alpha \! \approx \! 0.1$eV$\AA$ corresponding to the length scale $\frac{\hbar^2}{m^*\alpha}\!\! \sim \!\! 100$nm and $V_x\!\!\sim\!\! 1$K, $\Delta_0\!\!\sim\!\! 1$K, the critical current $I_c\!\sim\! 10$nA and $|\!L^{(\rm min)}_J\!(\varphi)| \!\!\sim\!\! 10\!-\!100$nH. The Josephson inductance $L_J(\varphi)$ can be probed by small-amplitude phase oscillations with the frequency $\omega$ satisfying $\omega\! \ll\! \Delta E\!\sim \! 10$GHz (for adiabatic approximation to hold)~\cite{note} and $\omega\! \gg \!1/\tau$ (to resolve current fluctuations). Thus, this experimental technique can be used to 
distinguish the Andreev spectrum in TP distinct phases and observe the phase transition we predict.

This work is supported by DARPA-QuEST and JQI-NSF-PFC. 


\end{document}